\pdfoutput=1

\documentclass[11pt]{article}
\usepackage{listings}
\usepackage{amsmath}
\usepackage{ACL2023}
\usepackage{graphicx}
\usepackage{booktabs}
\usepackage{subcaption}
\usepackage{amsmath}
\usepackage{algorithm}
\usepackage{algpseudocode}
\usepackage{latexsym}
\usepackage{makecell}
\usepackage{textcomp}
\usepackage{tabularx} 
\usepackage{times}
\usepackage{latexsym}
\usepackage{multirow}
\usepackage[T1]{fontenc}

\usepackage[utf8]{inputenc}
\usepackage{float}

\usepackage{microtype}

\usepackage{inconsolata}

\title{When LLMs Meet Acoustic Landmarks: An Efficient Approach to Integrate Speech into Large Language Models for Depression Detection}

\author{Xiangyu Zhang$^{1}$, Hexin Liu$^{2}$, Kaishuai Xu$^{3}$,\\ \textbf{Qiquan Zhang}$^{1}$, \textbf{Daijiao Liu}$^{1}$, \textbf{Beena Ahmed}$^{1}$, \textbf{Julien Epps}$^{1}$\\
The University of New South Wales$^1$\\
Nanyang Technological University$^{2}$
The Hong Kong Polytechnic University$^{3}$
}

\begin{document}
\maketitle

\begin{abstract}
Depression is a critical concern in global mental health, prompting extensive research into AI-based detection methods. Among various AI technologies, Large Language Models (LLMs) stand out for their versatility in mental healthcare applications. However, their primary limitation arises from their exclusive dependence on textual input, which constrains their overall capabilities. Furthermore, the utilization of LLMs in identifying and analyzing depressive states is still relatively untapped. In this paper, we present an innovative approach to integrating acoustic speech information into the LLMs framework for multimodal depression detection. We investigate an efficient method for depression detection by integrating speech signals into LLMs utilizing Acoustic Landmarks. By incorporating acoustic landmarks, which are specific to the pronunciation of spoken words, our method adds critical dimensions to text transcripts. This integration also provides insights into the unique speech patterns of individuals, revealing the potential mental states of individuals. Evaluations of the proposed approach on the DAIC-WOZ dataset reveal state-of-the-art results when compared with existing Audio-Text baselines. In addition, this approach is not only valuable for the detection of depression but also represents a new perspective in enhancing the ability of LLMs to comprehend and process speech signals. 
\end{abstract}

\begin{figure}[ht]
    \centering
    \includegraphics[width=0.45\textwidth]{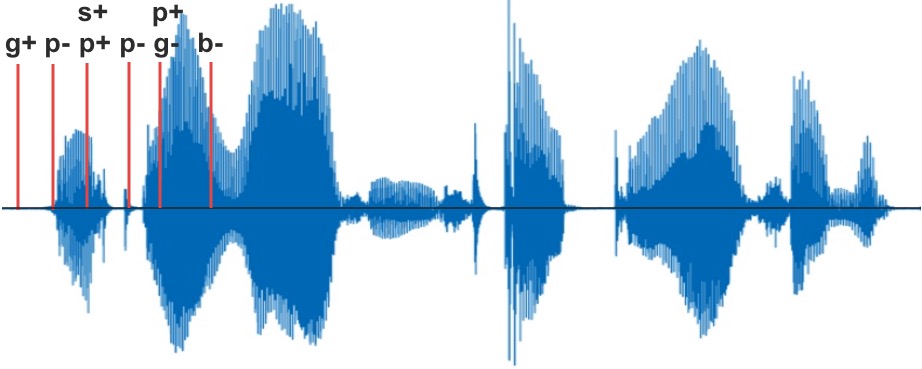}
    \caption{Example of Acoustic Landmark (2-gram concat landmark (g+p-), (s+p+), (p+,p-), ..., (g-b-)), Landmarks are extracted from abrupt changes in the speech signal. They can discretize speech into a series of tokens that possess linguistic significance.}
    \label{fig:landmark}
\end{figure}

\section{Introduction}
Depression, a common mental disorder affecting 10-15\% of the global population, is characterized by persistent low mood, loss of interest, and lack of energy, making it a prevalent and costly illness~\cite{walker2018prevalence}. Given the time-consuming, expensive, and sometimes ineffective nature of traditional depression treatment methods, a growing number of researchers are turning their attention to developing automated depression detection systems. Concurrently, Large language models (LLMs) have recently demonstrated remarkable success across a variety of tasks~\cite{chowdhery2023palm,touvron2023llama}. 
These large language models have been applied to various healthcare issues, including general surgery~\cite{oh2023chatgpt}, dementia diagnosis~\cite{wang2023can}, and gastroenterology~\cite{lahat2023evaluating} and achieved excellent results. However, their main limitation stems from their sole reliance on textual input, which limits their full potential. Simultaneously, the use of Large Language Models (LLMs) in depression detection remains largely unexplored. In particular, there has been no effort to integrate speech—despite growing evidence that speech signals can reveal indicators of depression~\cite{wu2023self, huang2019investigation}—into these LLMs, an advancement that could greatly improve their effectiveness in identifying depression ~\cite{zheng2023two}.

\begin{figure*}[ht]
    \centering
    \includegraphics[width=1\textwidth]{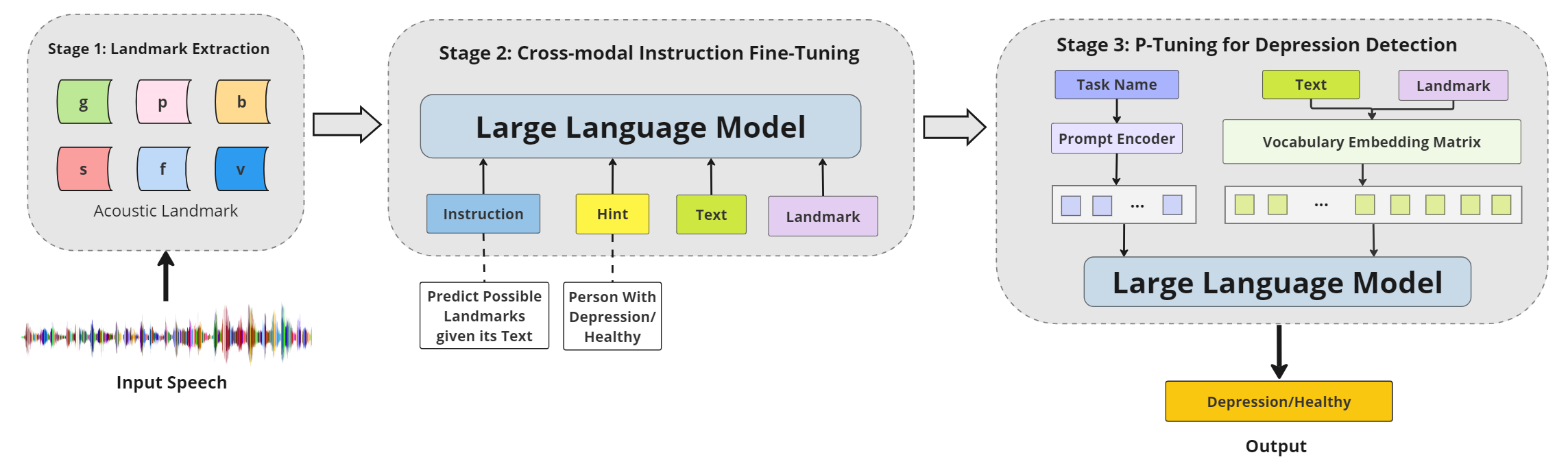}
    \caption{Overview of LLM-Landmark Depression Detection Pipeline, broadly categorized into three stages: landmark detection (on the left), cross-modal instruction fine-tuning (in the middle), and P-tuning for depression detection (on the right).} \vspace{-5mm}
    \label{fig:overview}
\end{figure*}
One of the key approaches to incorporating speech signals into LLMs is through the discretization of speech. However, the current landscape of speech discretization, heavily reliant on deep learning techniques~\cite{zeghidour2021soundstream,defossez2022high}, faces significant challenges due to its considerable GPU memory requirements. This is particularly problematic in the field of depression detection, where data often consists of lengthy conversations~\cite{devault2014simsensei}. The need for completed conversations is vital for accurate depression detection~\cite{wu2023self,sun2022tensorformer}, rendering the existing deep learning-based methods impractical for such applications. For this purpose, it is necessary to find an efficient approach that allows for the discretization of speech with reduced GPU memory usage.

Acoustic landmarks represent event markers intricately linked with the articulation of speech, forming a concise alternative framework for speech processing~\cite{liu1996landmark,stevens2002toward}. This approach emphasizes the analysis of abrupt acoustic changes at the subsegmental level, thereby providing a succinct and precise phonetic description of language.  These landmarks, characterized by their binary values, establish a minimal yet effective set for differentiating each language segment from others.  They maintain a direct and significant relationship with acoustic properties and articulation (including individual pronunciation), ensuring discernibility despite unwanted variability introduced by diverse hardware and environmental backgrounds~\cite{huang2018depression,huang2019speech}. Their discrete nature not only allows for efficient integration into large language models but also offers a viable alternative for understanding speech signals in depression detection, bypassing the limitations of current deep learning-based techniques. This innovative approach promises a more feasible and resource-efficient pathway for analyzing complex speech patterns in mental health diagnostics.

In this paper, we introduce a novel multimodal approach to depression detection, utilizing a combination of acoustic landmarks and large language models. We investigate the properties of large language models at various stages and under different conditions after integrating landmark-based speech information. We investigate how LLMs learn speech landmarks and assess the impact of conversational fine-tuning on the performance of LLMs in tasks related to depression detection.

In summary, our contributions include the following:
\begin{itemize}
    \item To the best of our knowledge, this is the first study to apply LLMs to \textbf{multimodal} depression detection and the inaugural effort to integrate speech information into LLMs for this purpose. We proposed a new baseline for the application of LLMs in the field of automatic depression detection.
    \item Compared with prior baseline audio-text methods~\cite{wu2023self}, our approach not only achieved SOTA performance but also involved a comprehensive analysis of the properties of LLMs post the integration of landmarks.
    \item Unlike previous deep learning-based methods for aiding LLMs in understanding speech, we explored a new, more efficient approach to enable LLMs to process speech signals. This novel method opens up a potentially groundbreaking direction for enhancing LLMs' comprehension of speech.
\end{itemize}

\begin{figure}[ht]
    \centering
    \includegraphics[width=0.5\textwidth]{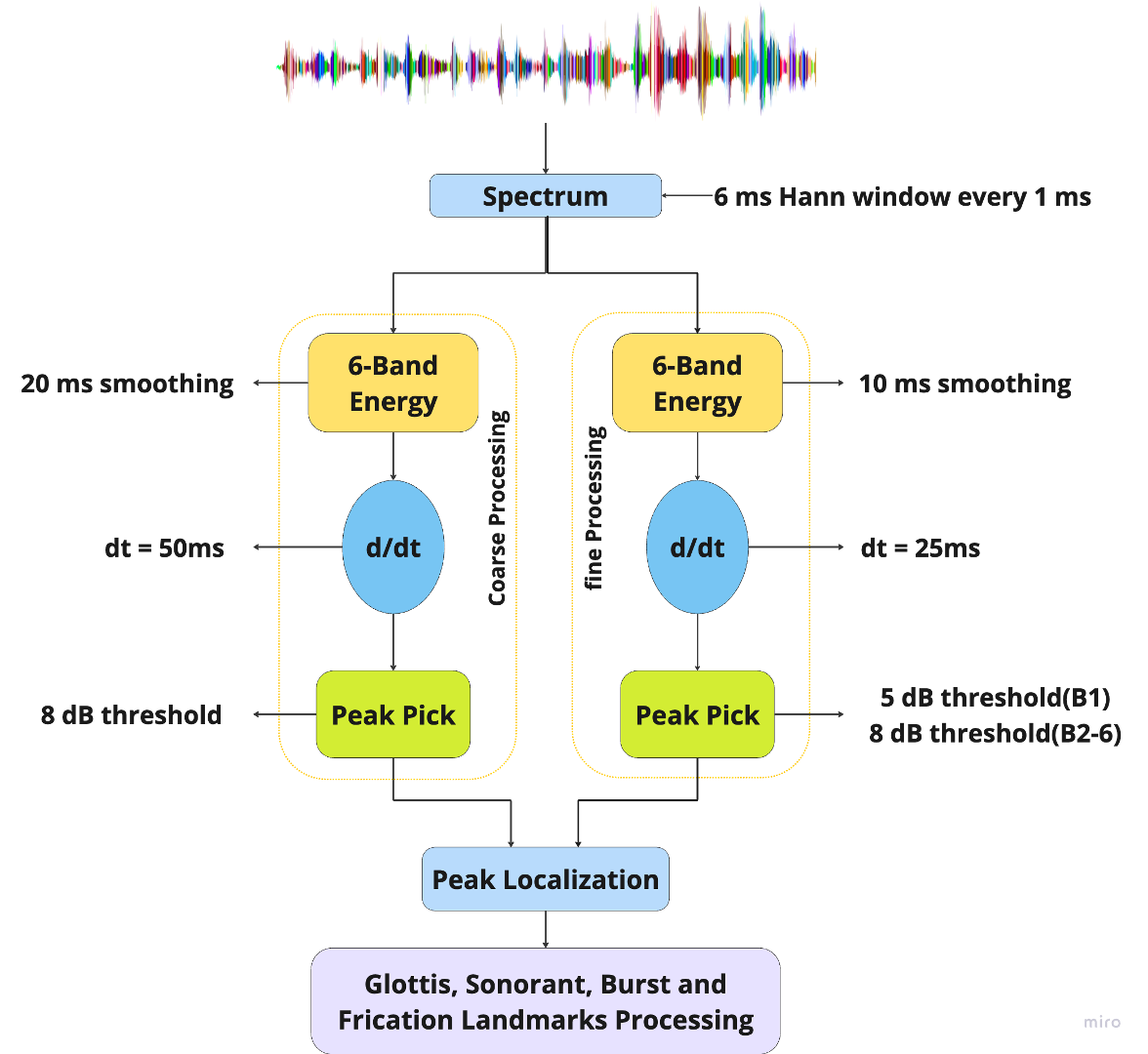}
    \caption{Landmark Detection Filter}\vspace{-5mm}
    \label{fig:filter}
\end{figure}

\section{Related Work}

\subsection{Large Language Models}
Large language models have achieved success in natural language processing and have been extended to encompass computer vision and speech signal processing~\cite{gpt3, touvron2023llama,li2023pqlm,liu2024aligning}. However, there is a significant gap in research aimed at enabling LLMs to comprehend speech efficiently.

Parameter-efficient fine-tuning refers to selectively updating a small subset of the model's parameters or adding lightweight trainable layers, to customize the model for specific tasks or domains with reduced computational overhead. Existing works employed low-rank adaptation~(LoRA) to fine-tune LLM efficiently. LoRA reduces computational complexity by freezing the pre-trained LLM and injecting trainable rank decomposition matrices A and B into its transformer-based layers~\cite{hu2022lora}. The forward pass is subsequently defined as the linear combination of those from the pre-trained model and from the trained decomposed matrices A and B.

\subsection{Acoustic Landmarks}
The concept of acoustic landmarks originally stems from research on distinctive features~\cite{garvin1953preliminaries,zhang2024auto}. Some researchers posit that for certain phonetic contrasts, a listener relies on acoustic landmarks to gather the necessary acoustic cues for deciphering the underlying distinctive features~\cite{liu1996landmark}. This perspective highlights the importance of these landmarks in the auditory processing and interpretation of speech. Subsequent research has utilized acoustic landmarks for applications in speech recognition~\cite{liu1996landmark,he2019ctc} as well as in addressing mental health-related problems~\cite{huang2018depression, huang2019investigation}. Although different scholars have slightly varied definitions of acoustic landmarks, Joel and colleagues~\cite{boyce2012speechmark} expanded upon Liu's paper~\cite{liu1996landmark} by releasing a MATLAB version of a landmark detection toolkit, which has become the most widely used version of landmark technology.

\subsection{Automatic Depression Detection}
The use of AI technology for depression detection has been developing for many years. Some researchers ~\cite{cummins2011investigation, huang2018depression,huang2019investigation} have utilized traditional methods such as Support Vector Machines (SVMs)~\cite{noble2006support} for depression detection. With the advancement of deep learning technologies~\cite{gulati2020conformer,zhang2024mamba}, an increasing number of researchers have been experimenting with deep learning approaches for depression detection. Zhao and others have explored the use of transformer models for processing speech inputs in depression detection~\cite{zhao2020hierarchical}. Shen and colleagues have employed BI-LSTM architectures, combining text and speech for this purpose~\cite{shen2022automatic}. Further extending these techniques, Wu~\cite{wu2023self} utilized speech self-supervised models~\cite{chen2022wavlm,hsu2021hubert,efficient_lid} and integrated them with RoBERTa~\cite{liu2019roberta} for a more comprehensive text-audio multimodal approach to depression detection.

\section{Methodology}
\subsection{Overview}
Our methodology, detailed in Figure \ref{fig:overview}, encompasses a three-step training process. The first phase involves extracting acoustic landmarks from speech and conducting an array of data processing operations. Subsequently, in the Cross-modal Instruction Fine-Tuning phase, we engage the LLM in learning the nuances and characteristics of acoustic landmarks. The culminating phase is the P-Tuning process, wherein the LLM is meticulously trained to apply its understanding to diagnose depression. 
\subsection{Landmarks Extraction and Data Preprocessing}
\subsubsection{Landmarks Extraction}
Figure \ref{fig:landmark} illustrates an example of acoustic landmarks, where speech signals are discretized into a series of symbols that carry linguistic relevance. Table \ref{landmark_table}
details the specific acoustic landmarks utilized in our study. Diverging from Liu's paper~\cite{liu1996landmark}, our research also pays attention to frication, voice frication, and periodicity.








\begin{table}[ht]
\centering
\begin{tabular}{l| p{0.65\linewidth}}
\hline
\hline

\textbf{Landmark} & \textbf{Description} \\
\hline
\hline
\multirow{2}{*}{g} & vibration of vocal folds start (+) or end (–) \\ \hline
\multirow{3}{*}{b} & onset (+) or offset (–) of existence of turbulent noise during obstruent regions \\ \hline
\multirow{2}{*}{s} & releases (+) or closures (–) of a nasal \\ \hline
\multirow{2}{*}{v} & voiced frication onset (+) or offset (–) \\ \hline
p & periodicity start (+) or end (–) \\ \hline
f & frication onset (+) or offset (–) \\
\hline 
\hline
\end{tabular}
\caption{Description of the six landmarks investigated.}
\vspace{-5mm}
\label{landmark_table}
\end{table}

Our method primarily draws inspiration from Joel's~\cite {boyce2012speechmark} and Liu's~\cite{liu1996landmark} work. However, since they have not open-sourced their code, many of their approach's details remain unknown. In the following section, We introduce our Python-based landmark detection algorithm, developed to address these gaps and to adapt the conceptual framework to our specific requirements. Initially, the spectrogram is divided into six frequency bands. Landmarks are identified through energy changes within these six bands, using a two-pass strategy. Different landmarks are determined by either a single band or a combination of multiple bands~\cite{liu1996landmark}. This approach is visually represented by the two parallel branches emanating from the spectrogram block in Figure \ref{fig:filter}. 

The detection algorithm for \textbf{Glottal (g)}, \textbf{Burst (b)}, and \textbf{Syllabic (s)} landmarks is fundamentally aligned with Liu's approach \cite{liu1996landmark}. However, diverging from Liu's method, we employ 5dB and 8dB as threshold values because of different smoothing methods between Python and Matlab. Additionally, considering that the opening and closing of the glottis occur in pairs, We implemented dynamic programming to ensure that g landmarks appear in pairs, thus enhancing the physiological accuracy of our detection. 

Our methodology for identifying \textbf{f+} and \textbf{v+} landmarks involves detecting a 6 dB power increase in at least three high-frequency bands (bands 4-6), and a power decrease in low-frequency bands (bands 2 and 3). For \textbf{f-} and \textbf{v-}, the criteria are reversed: a 6 dB power decrease in the same high-frequency bands and a power increase in the low-frequency bands. The distinguishing factor here is that frication landmarks are detected within unvoiced segments (b landmark), while voiced frication landmarks are sought in voiced segments (s landmark). 

Regarding the detection of the \textbf{periodicity (p)}  landmarks, we perform autocorrelation calculations on the audio frame to identify repetitive or periodic patterns in the data. For a detailed description of our landmark detection algorithm, please refer to Appendix \ref{appendix:A}.
\subsubsection{Data Augmentation and Processing}
Depression assessments are commonly conducted through clinical interviews, with each session receiving a singular label. This labeling method, when applied to a given dataset size, leads to fewer samples in datasets compared with the much larger number of utterances and frames typically encountered in other speech-related tasks. As a result, the speech depression detection task faces a notable challenge of data scarcity. Moreover, the issue of data imbalance is particularly acute in the dataset, as instances of healthy (positive cases) are significantly outnumbered by depression (negative) cases. We adopted Wu's approach~\cite{wu2023self} of augmenting the training set through sub-dialogue shuffling. Sub-dialogue shuffling involves sampling a sub-dialogue \( x_{s:e} \) from each complete dialogue \( x_{1:T} \), where \( s \) and \( e \) represent the randomly selected start and end utterance indexes, respectively.

This technique allowed us to balance the number of positive and negative samples effectively, while substantially increasing the dataset size. Differing from Wu's method, our use of landmarks in speech processing enables the use of longer sub-dialogues for training purposes. To ensure a fair comparison, we maintained the same data size (same sub-dialogue sampling number \textit{M}=1000) as Wu's approach. For a detailed description of the algorithm, please refer to Appendix \ref{appendix:B}.

Previous research has indicated that the patterns in which landmarks appear are more valuable than the individual landmarks themselves~\cite{huang2019investigation}. Therefore, as shown in Figure \ref{fig:landmark}, we combined landmarks, treating every two consecutive landmarks as a single unit. This approach not only better represents the patterns of landmarks but also effectively reduces the length of the landmark sequence in each sample.

\subsection{Hint Cross-modal Instruction Fine-Tuning}
Since LLMs inherently lack exposure to acoustic landmarks, our initial step involves devising a method to teach the LLM what acoustic landmarks are. This foundational training is crucial for enabling the models to interpret and utilize acoustic landmark data effectively. 

\begin{table*}[ht]
    \centering
    \setlength{\abovetopsep}{0pt}
    \setlength\belowbottomsep{0pt} 
    \setlength\aboverulesep{0pt} 
    \setlength\belowrulesep{0pt}
    \setlength{\tabcolsep}{3.5pt}
    \begin{tabular}{l|cccccc}
    \toprule[1.5pt]
         Method/ Model & Llama2-7B & Llama2-7B Chat & Llama2-13B & Llama2-13B Chat & GPT3.5 & GPT4 \\
         \hline
         Text Only&  0.578 & 0.488 & 0.636 & 0.545 & 0.545 & 0.571\\
         Landmark Only& 0.521 & 0.434 & 0.559 & 0.538& - & - \\
         Text + Landmark & 0.545 & 0.500 & 0.695 & 0.666 & - & -\\
    \toprule[1.5pt]
    \end{tabular}
    \caption{F1 scores for the different LLM models, We test all Llama2 models for 7B and 13B, also test on GPT.}\vspace{-5mm}
    \label{Main Result}
\end{table*}

As depicted in the middle section of Figure \ref{fig:overview}, our task involves providing an LLM with instructions to predict potential acoustic landmarks based on text. This method serves a dual purpose: it enables the LLM to learn about acoustic landmarks, and it also aligns speech (landmarks) and text modalities using paired data. We adopt LoRA~\cite{hu2022lora} by incorporating low-rank matrices into the Query and Key matrices of the self-attention layer, facilitating efficient adaptation and fine-tuning. Additionally, we resize the embedding layer of the LLMs to add the merged landmarks to the vocabulary. During the training process, both the \textbf{embedding layer}, \textbf{linear head} and the \textbf{LoRA matrices} are actively trained to integrate these new elements effectively. The training objective is to minimize the negative log-likelihood, and the loss calculation applies to all samples (including the prefix), which can be formulated as:
\begin{equation}
    \mathcal{L}(M|C) = - \sum_{j=1}^{x} \sum_{i=1}^{y_j} \log P(s_{i,j}|s_{<i,j},M),
\end{equation}
where \( x \) is the number of samples in dataset \( C \), \( y_j \) is the text and corresponding landmarks in sample \( S \), and \( M \) denotes the large language model that we have fine-tuned.

Additionally, during dataset construction, we incorporate hints for the LLM. For example, when data are sourced from a patient with depression, we include a hint indicating their origin from a depressed patient.
Experimentally, we found this method of data construction to be crucial, which also supports our hypothesis that \textbf{the acoustic landmarks from individuals with depression differ from those of healthy individuals}. For detailed template construction, please refer to Appendix \ref{appendix:C}.

\subsection{P-Tuning for Depression Detection}
In the previous stage, we trained the LLMs to understand what landmarks are. Following this, we employ P-tuning~\cite{LIU2023} to enable the LLMs to integrate text and landmarks for depression detection. We replace the lm head layer with the classification layer. The training objective is to minimize cross-entropy for classification, which can be formulated as 
\begin{equation}
    \mathcal{L}= -\sum_{c=1}^{C} y_{o,c} \log(p_{o,c}),
\end{equation}
where \( C \) is the number of classes. \( y_{o,c} \) is an indicator variable that is 1 if the observation \( o \) belongs to class \( c \) and 0 otherwise. \( p_{o,c} \) is the predicted probability of observation \( o \) belonging to class \( c \).
We also compared instruction tuning using LoRA with P-tuning and discovered that \textbf{manually constructed templates are not well-suited for depression classification tasks}. Furthermore, we observed a performance improvement when applying LoRA matrices across all layers of Llama2.

\subsection{Decision Making}
In the previous study by ~\cite{wu2023self}, they achieved state-of-the-art (SOTA) results through an ensemble approach, combining WavLM, WavLM pre-trained on emotional recognition tasks, and the combined result of RoBERTa and WavLM. Adopting a similar strategy, we fine-tune three distinct LlaMA2 (Text + Landmark) models, each with different data volumes (different numbers of sub-dialogue M(900, 1000, 1100)), and used them for ensemble voting.

\section{Experiments}
\subsection{Experimental Setup}

\textbf{Dataset}. The DAIC-WOZ dataset~\cite{devault2014simsensei}, recognized as a standard for depression detection, includes 189 clinical interview recordings between interviewers and patients. In its training subset, 30 of the total 107 interviews are labelled as depressed, while the development subset contains 12 depressed instances out of 35 interviews. Consistently with previous studies~\cite{gong2017topic,shen2022automatic,wu2022climate,wu2023self}, we report our results on the development subset.
\begin{table}[ht]
\centering
  \normalsize
  \setlength{\abovetopsep}{0pt}
    \setlength\belowbottomsep{0pt} 
    \setlength\aboverulesep{0pt} 
    \setlength\belowrulesep{0pt}
\renewcommand{\arraystretch}{1.1}
\resizebox{\columnwidth}{!}{%
\begin{tabular}{l|l|c|c}
\toprule[1.5pt]
                  \textbf{Methods} & \textbf{Model} & \textbf{F1} & \textbf{Ensemble} \\ \hline
                  
\multirow{3}{*}{\makecell[c]{\textbf{Previous SOTA} \\ \cite{wu2023self}}}  & WavLM + RoBERTa& 0.648  & \multirow{3}{*}{0.829}\\
& WavLM Layer 8                         & 0.700    &   \\
& WavLM Layer 10 & 0.720 & \\ \hline
\multirow{3}{*}{\makecell[c]{\textbf{Text+Landmark} \\ (Our)}}    & Llama2 ($M\!=\!900$)   & 0.636 & \multirow{3}{*}{\textbf{0.833}}      \\ 
                  & Llama2 ($M\!=\!1000$)         & 0.695       &   \\
                  & Llama2 ($M\!=\!1100$)         & 0.719       &    \\ 
                  \hline
\toprule[1.5pt]

\end{tabular}
}
\caption{
A comparison of our proposed system with previous state-of-the-art (SOTA), where all ensemble outcomes(F1 Score) are derived from a majority vote. In the table, \textit{M} denotes the number of augmented sub-dialogues per dialogue in our data augmentation algorithm, while the previous SOTA used \textit{M}=1000 sub-dialogues.}\vspace{-5mm}
\label{Compare Result}
\end{table}
\newline
\textbf{Model Configurations}.
Our research utilizes Llama2-7B, Llama-7B Chat, Llama2-13B, and Llama2-13B Chat, conducted on a system equipped with 8 NVIDIA A100 80GB GPUs. Llama 2-Chat was optimized for engaging in two-way conversations. In the cross-modal instruction fine-tuning stage, We fine-tuned the model with 10 epochs with 128 batch sizes, 8 Lora ranks, 100 warmup steps, and a 1e-6 learning rate. In the depression detection stage, we fine-tuned the model with 8 epochs with 256 batch sizes, 30 virtual tokens, 256 encoder hidden sizes, and a 1e-6 learning rate. In both experiments, we used AdamW as an optimizer with the model parallel to fine-tune our model. In the ablation study stage, we used hyperparameter tuning following the Tree-structured Parzen Estimator (TPE) paradigm~\cite{bergstra_algorithms_2011}.

\subsection{Main Result: Performance of different LLMs in Depression Detection task}
\textbf{Depression Detection in Llama2}. Table \ref{Main Result} displays the F1 scores obtained by Llama2 in depression detection across different scenarios. Additionally, we conducted a comparison of our findings with the results obtained from GPT-3.5 and GPT-4, focusing solely on their performance in the text modality.  It is crucial to highlight that we did not fine-tune GPT-3 or GPT-4 for our purposes. Rather, we employed carefully crafted prompts(see appendix~\ref{appendix:D}), allowing the GPT models to assess whether a particular sample was from a patient with depression. 

For the 'landmark only' and 'landmark + text' results, the process involved first undergoing hint cross-modal instruction fine-tuning and then employing P-tuning for depression detection. The objective was to equip the LLMs with a preliminary understanding of landmarks before advancing to the diagnostic stage for depression.

The experimental results reveal that when LLMs solely use the text modality for depression detection, the performance of all models, including notably powerful ones like GPT-3.5 and GPT-4, which excel in many tasks, is not particularly impressive and remains somewhat unsatisfactory. We attribute the subpar performance to two main factors. First is the \textbf{inherent limitation of the text modality in conveying emotional information}. For instance, consider the sentence, "It's raining today." While some may find this statement positive, others might feel the opposite. It's challenging to discern the emotional nuances from the text alone, but with audio information, we could accurately capture the emotional context of the statement. Secondly, \textbf{the issue lies with the data itself}. Labels are only available at the document level, and data are scarce (currently, there are no larger public datasets available for multimodal depression detection). This limitation in data granularity and volume significantly hinders the model's ability to accurately detect depression.

The introduction of landmarks led to enhanced performance across all models, affirming the effectiveness of our method in integrating landmarks. Landmarks can represent some of the acoustic information due to affective variation, providing additional information that assists LLMs in detecting depression. Nonetheless, the efficacy of using landmarks in isolation for depression detection was found to be suboptimal. Drawing on past research, we believe this is due to the fact that even after cross-modal instruction fine-tuning, relying solely on information from other modalities (such as audio or visual) could potentially impair the stability of LLMs~\cite{zhang2023llama,li2023prompting}. When we combined multiple Llama2 models that had integrated both text and landmark information for depression detection, we achieved SOTA results as shown in table \ref{Compare Result}. Furthermore, as indicated in Table \ref{Compare Result}, there is a gradual improvement in Llama2's performance in depression detection tasks as the number of sub-dialogues per dialogue increases. This observation further emphasizes the crucial role that data quantity plays in the effectiveness of depression detection tasks.

\begin{figure*}[ht]
\centering

\begin{subfigure}[b]{0.23\textwidth}
    \includegraphics[width=\textwidth]{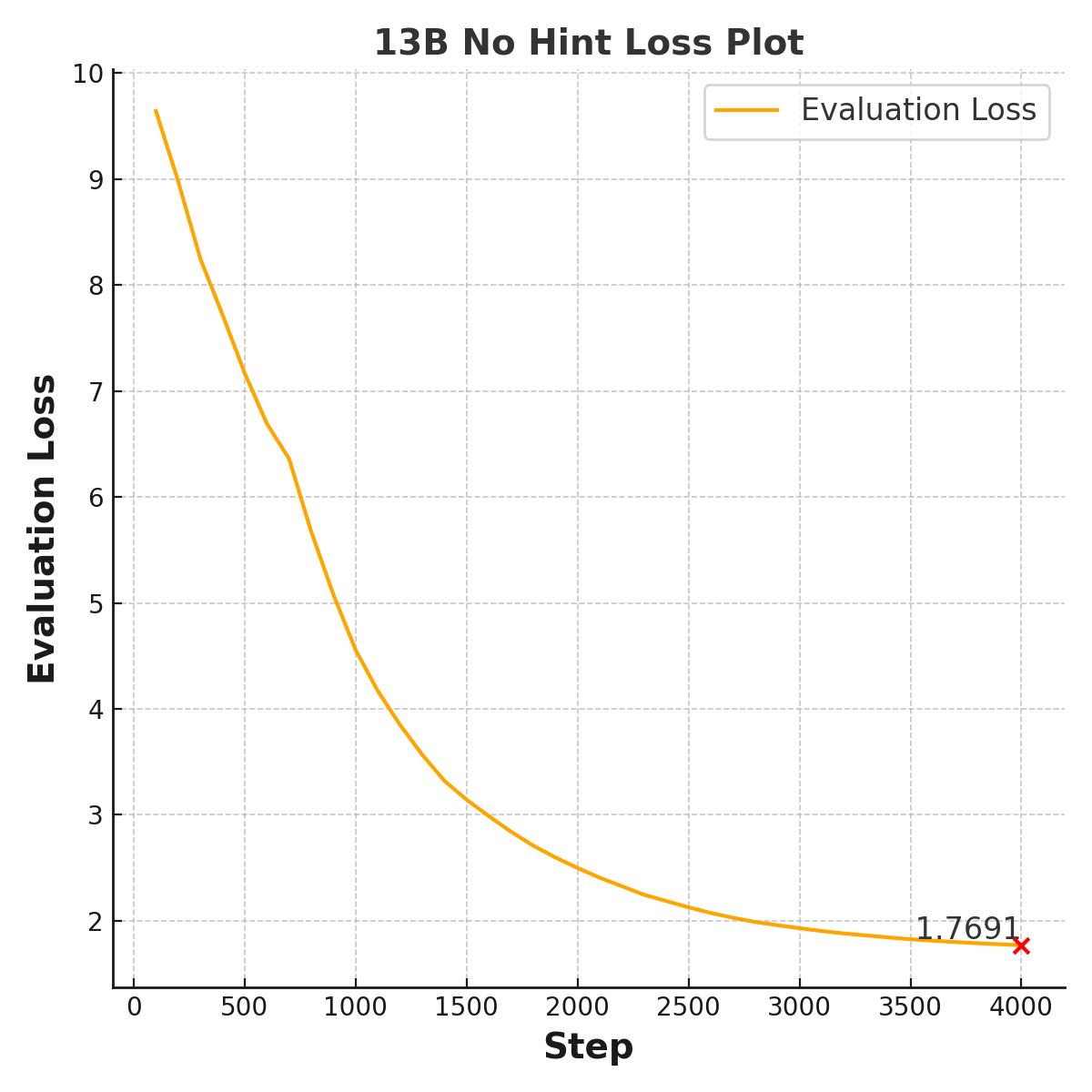}
    \caption{13B No Hint}
    \label{fig:13b_no_hint}
\end{subfigure}
\hfill
\begin{subfigure}[b]{0.23\textwidth}
    \includegraphics[width=\textwidth]{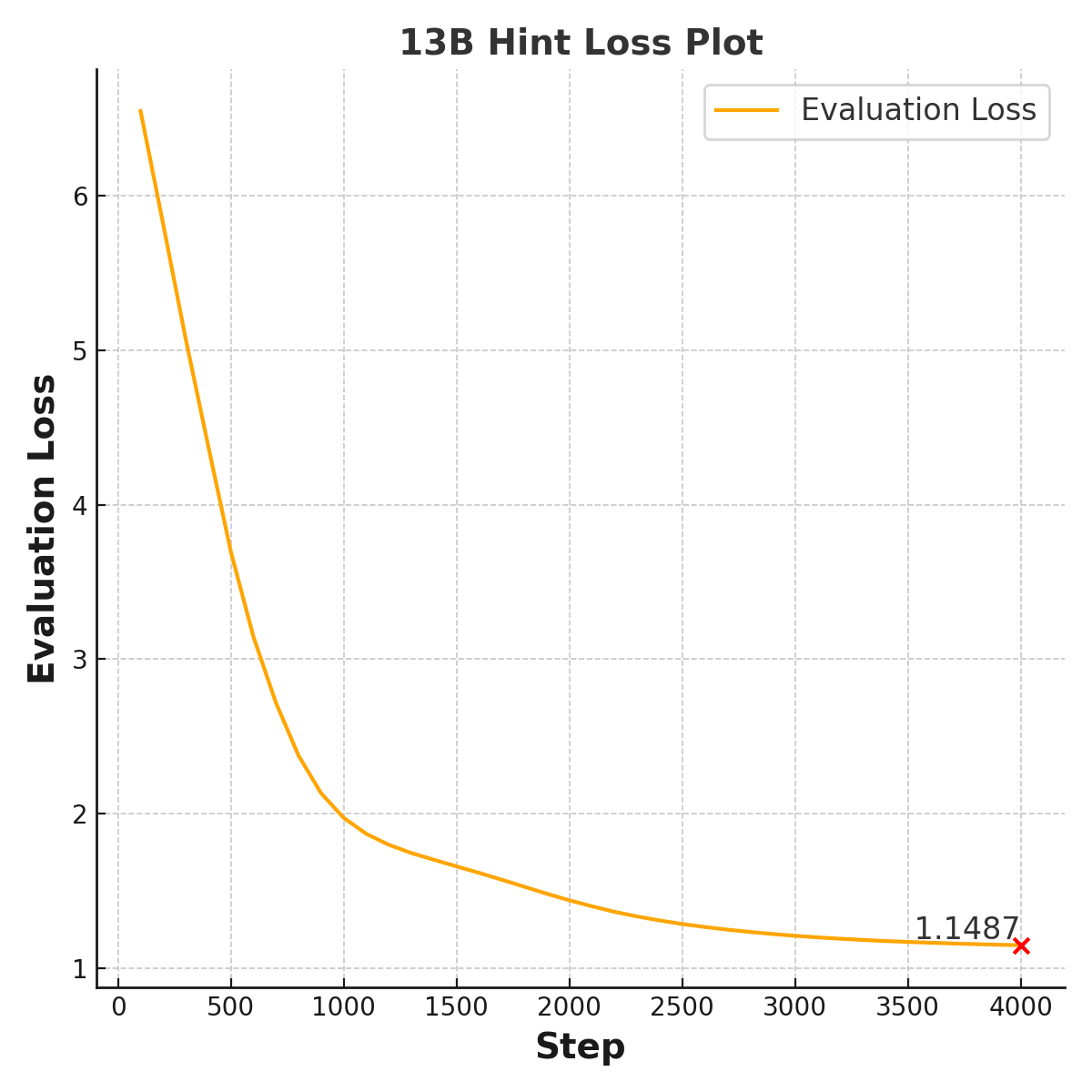}
    \caption{13B Hint}
    \label{fig:13b_hint}
\end{subfigure}
\hfill
\begin{subfigure}[b]{0.23\textwidth}
    \includegraphics[width=\textwidth]{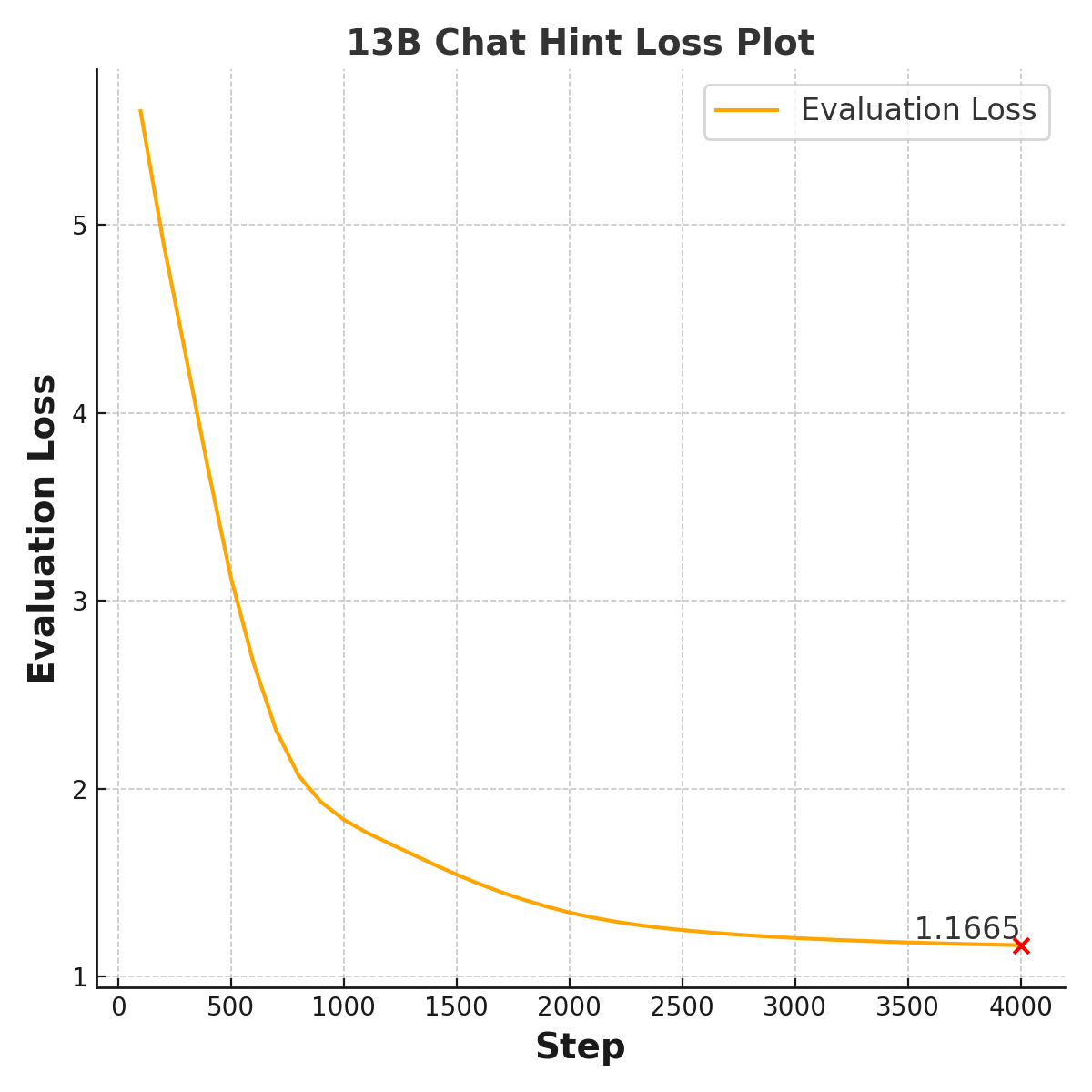}
    \caption{13B Chat Hint}
    \label{fig:13b_chat_hint}
\end{subfigure}
\hfill
\begin{subfigure}[b]{0.23\textwidth}
    \includegraphics[width=\textwidth]{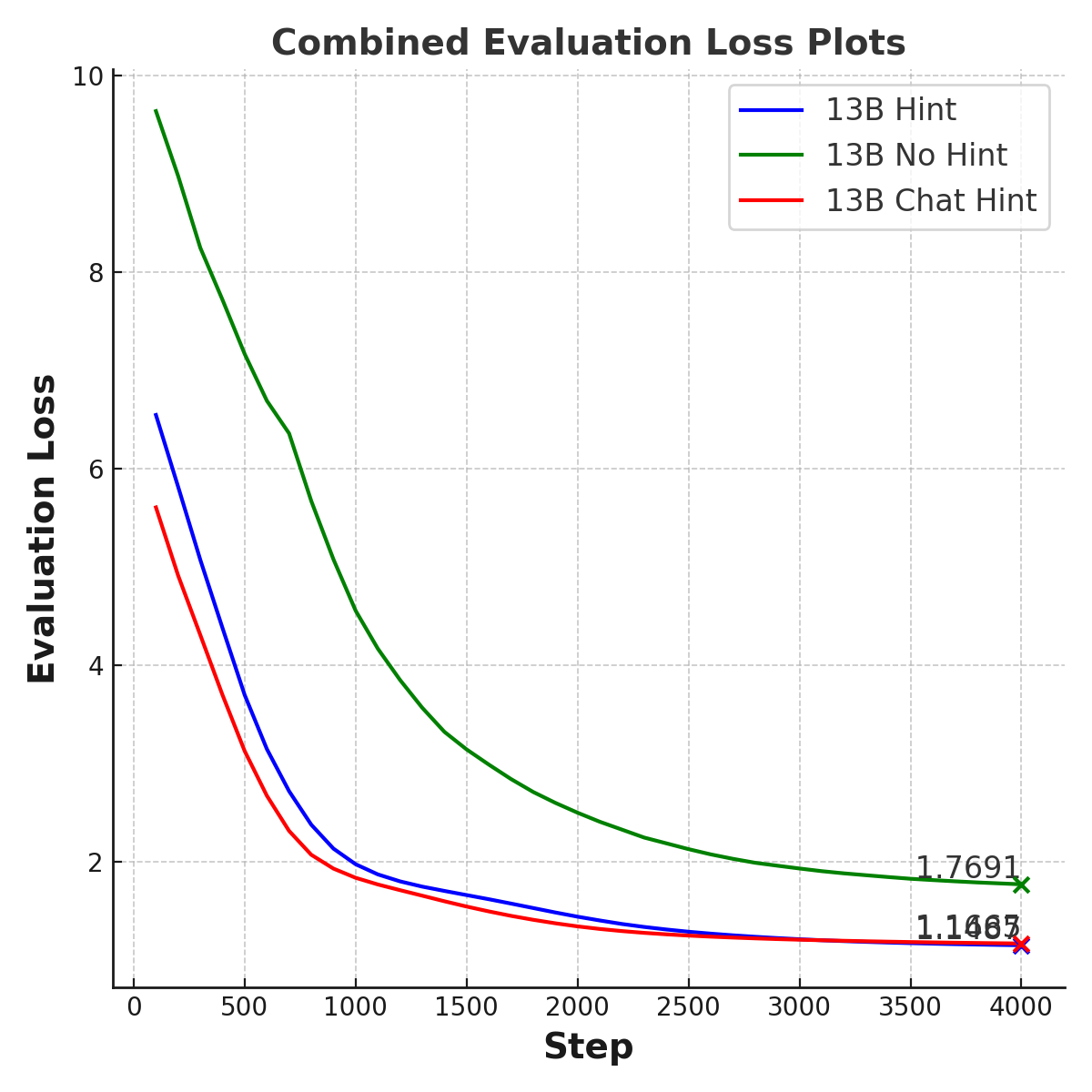}
    \caption{Three Comparison}
    \label{fig:combined}
\end{subfigure}

\caption{Evaluation loss for different configurations up to 4000 steps.}\vspace{-3mm}
\label{fig:eval_loss}
\end{figure*}

\vspace{-3mm}
\section{Ablation Study and Discussion}
\vspace{-3mm}
In this chapter, we conduct an empirical study to meticulously analyze and elucidate the characteristics of LLMs that we identified in the context of depression detection during our experiments.

\subsection{Effect of Hint in Cross-Modal Instruction Fine-Tuning}

During the Cross-Modal Instruction Fine-Tuning phase, we discovered that providing a hint to the LLMs is crucial. In other words, informing the LLMs whether the data sample originates from a patient with depression significantly impacts the training outcome. As evident from Figure~\ref{fig:eval_loss}, without a hint, the loss converged to around 1.76 (as shown in Figure~\ref{fig:13b_no_hint}). In contrast, with a hint, the loss consistently converged to near 1.1 (as depicted in Figures~\ref{fig:13b_hint} and~\ref{fig:13b_chat_hint}). Figure~\ref{fig:combined} offers a more vivid illustration of the substantial difference that the presence or absence of a hint makes to the model's performance in our empirical study. This phenomenon supports our previous conjecture that \textbf{individuals with depression and those who are healthy differ in their vocal expressions and that landmarks are capable of reflecting this characteristic}. Although the differences between Llama2 and Llama2 Chat are not substantial, it is still observable that, in this phase, Llama2 outperforms its Chat version. We will provide a more detailed discussion in the subsequent section.

\subsection{How LLMs Learn from Acoustic Landmarks}

\begin{figure*}[ht]
\centering
\begin{subfigure}[b]{0.24\textwidth}
    \includegraphics[width=\textwidth]{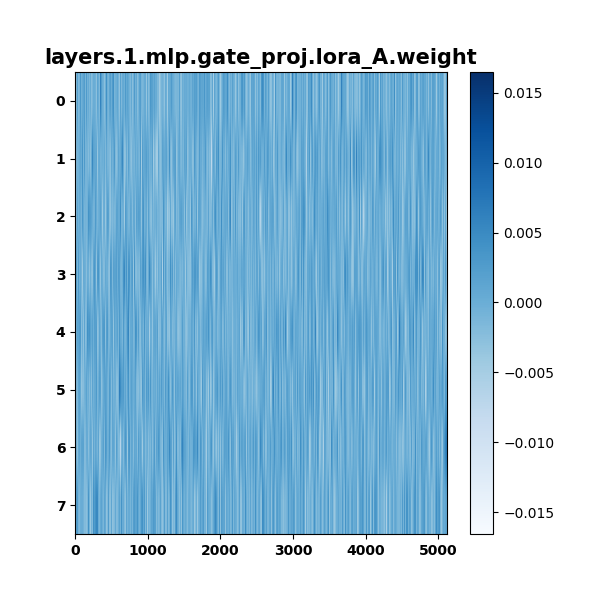}
    \caption{Top 1 Contribution Layer}
    \label{Top 1}
\end{subfigure}
\hfill
\begin{subfigure}[b]{0.24\textwidth}
    \includegraphics[width=\textwidth]{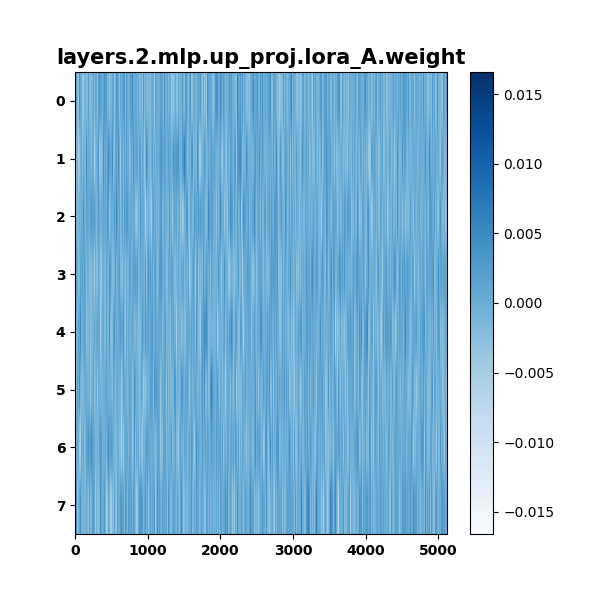}
    \caption{Top 2 Contribution Layer}
    \label{Top 2}
\end{subfigure}
\hfill
\begin{subfigure}[b]{0.24\textwidth}
    \includegraphics[width=\textwidth]{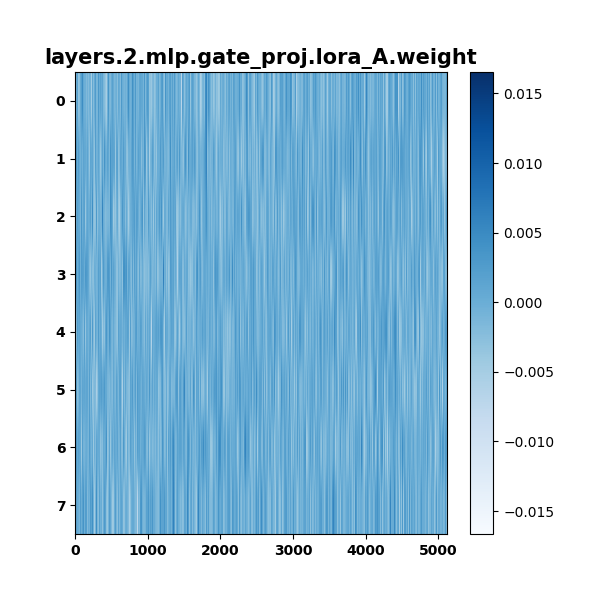}
    \caption{Top 3 Contribution Layer}
    \label{Top 3}
\end{subfigure}
\hfill
\begin{subfigure}[b]{0.24\textwidth}
    \includegraphics[width=\textwidth]{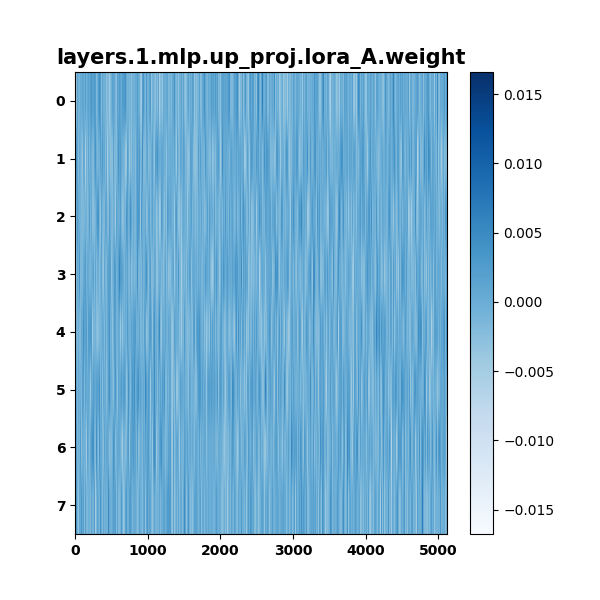}
    \caption{Top 4 Contribution Layer}
    \label{Top 4}
\end{subfigure}
\begin{subfigure}[b]{0.24\textwidth}
    \includegraphics[width=\textwidth]{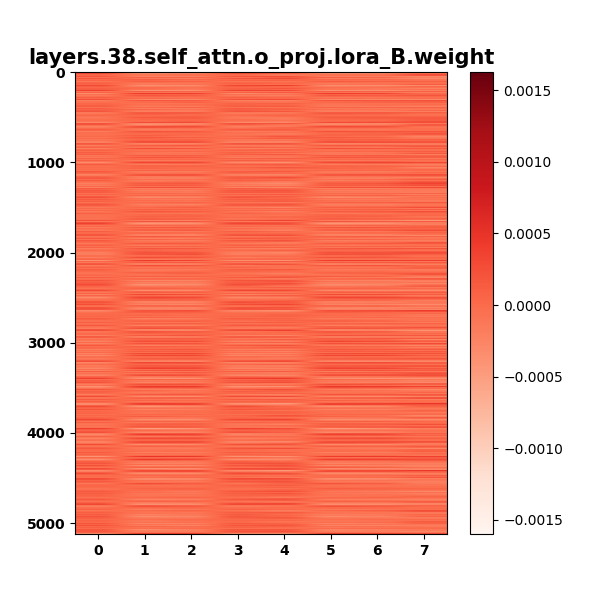}
    \caption{Bottom Layer 1 }
    \label{B1}
\end{subfigure}
\hfill
\begin{subfigure}[b]{0.24\textwidth}
    \includegraphics[width=\textwidth]{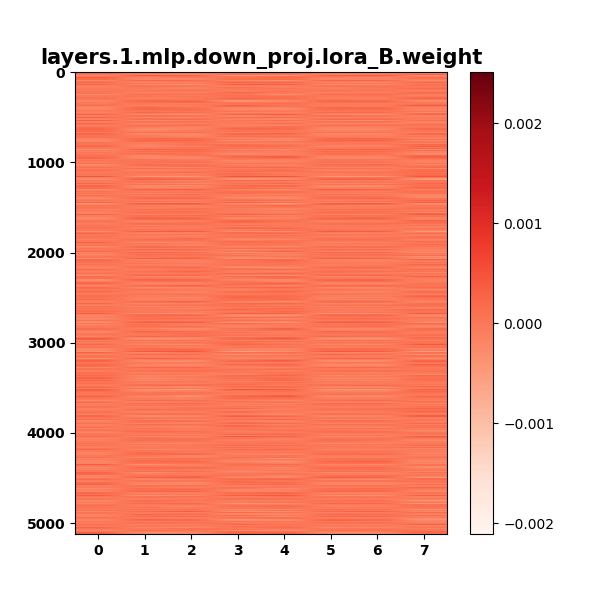}
    \caption{Bottom Layer 2}
    \label{B2}
\end{subfigure}
\hfill
\begin{subfigure}[b]{0.24\textwidth}
    \includegraphics[width=\textwidth]{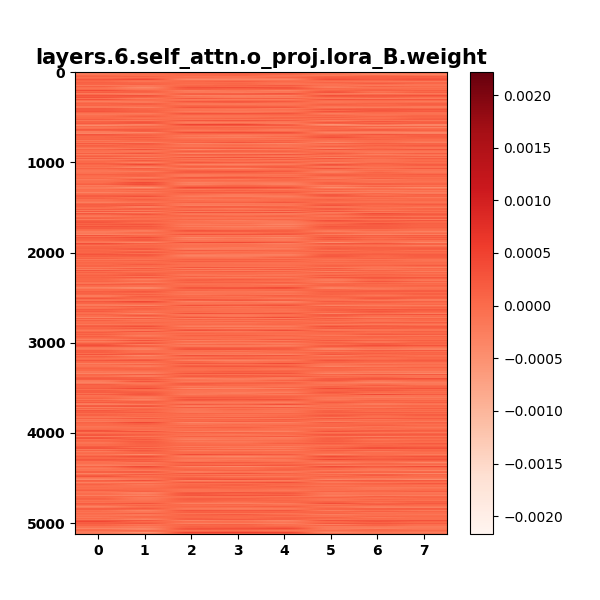}
    \caption{Bottom Layer 3}
    \label{B3}
\end{subfigure}
\hfill
\begin{subfigure}[b]{0.24\textwidth}
    \includegraphics[width=\textwidth]{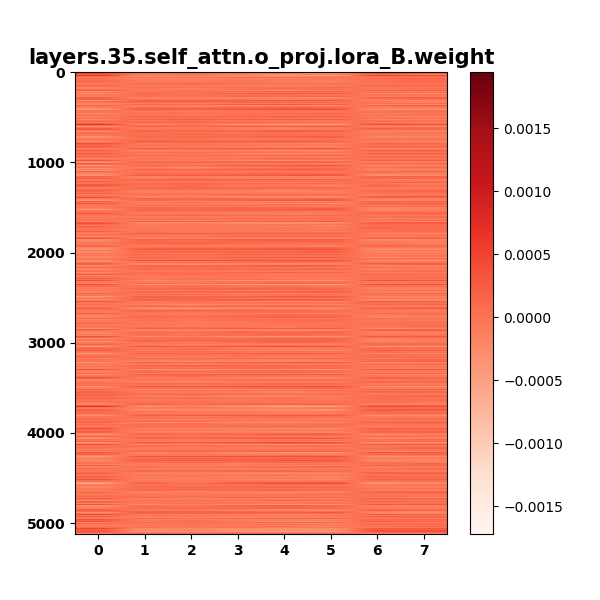}
    \caption{Bottom Layer 4}
    \label{B4}
\end{subfigure}
\caption{The top four images represent the LoRA matrices of the layers that contribute most significantly to the large language model's learning of landmarks. The bottom four images depict the LoRA matrices of the layers with the least contribution. As can be inferred from the graph's title, the feedforward layer is the primary contributor.}
\vspace{-6mm}
\label{contri matrix}
\end{figure*}

To further investigate how LLMs learn acoustic landmarks, we extended the application of LoRA beyond just the attention layers, applying it across all layers for comprehensive analysis~\cite{pu2023empirical,sun2023comparative,li2023quantitative,zhang2024rethinking}. To find the matrix with the greatest contribution, we first need to define the method for calculating the contribution of a matrix. We can approximately consider the changes in the LoRA matrix as indicative of its contribution to the task~\cite{he2021towards}. Therefore, we assess that the contribution of a matrix is calculated by summing the absolute values of all its elements, normalized by the total number of elements in the matrix. Suppose we have a set of LoRA matrices \( L_1, L_2, \ldots, L_n \), each matrix \( L_i \) being an \( a \times b \) matrix. Then, the contribution \( C_i \) of matrix \( L_i \) can be calculated using the formula:
\begin{equation}
     C_i = \frac{1}{ab} \sum_{j=1}^{a} \sum_{k=1}^{b} |L_i(j, k)|. 
\end{equation}
Here, \( |L_i(j, k)| \) represents the absolute value of the element in the \( j^{th} \) row and \( k^{th} \) column of matrix \( L_i \). After calculating the contribution value (C), we rank and select the ten matrices with the highest and the lowest contributions for further analysis. Figure~\ref{contri matrix} separately illustrates the four matrices with the greatest contributions and the four with the least. To validate the effectiveness of this method, we deactivated the five matrices with the smallest contributions and observed that this had no significant impact on our results.

Our analysis of the matrices revealed that LLMs primarily \textbf{learn landmarks through the feedforward network}, while the contribution of the LoRA matrices in the attention layers is quite minimal. This phenomenon is also observed when training LLMs to learn speech codecs~\cite{hao2023boosting}, suggesting that even though landmarks have inherent linguistic significance, LLMs tend to treat landmarks as abstract tensors, similar to speech codecs, during the learning process. Additionally, we observed that \textbf{layers closer to the beginning of the LLMs have a greater contribution} to learning landmarks. This could be because LLMs treat landmarks as new vocabulary items, leading to more updates in layers nearer to the embedding layer.

\subsection{Llama2 vs Llama2 Chat, and Generation vs Classification}
LlaMA2 models are uncensored and have not undergone instruction tuning or chat-tuning. In contrast, LlaMA2 Chat models are censored and have been chat-tuned, making them optimized for dialogue use cases~\cite{touvron2023llama}. When treating depression detection as a classification task, we tested LlaMA2 Chat and found that its performance, both during the Cross-modal Instruction Fine-Tuning stage and the depression detection phase, was inferior to that of LlaMA2. We hypothesize two potential reasons for this. The first is that the Chat version might not be suitable for classification tasks. The second, and our preferred explanation, is that the Chat version, having been adjusted, tends to avoid answering questions to mitigate ethical risks. To validate our hypothesis, we first reimagined the classification task as a generative task, where the LLMs diagnoses depression through dialogue responses. We tested this zero-shot scenario on GPT-3.5 and GPT-4. Additionally, we applied LoRA for instruction fine-tuning in various scenarios presented in Table~\ref{Main Result}, to observe how the models perform post-tuning. We observed that when treating depression detection as a generative task, neither LlaMA2 nor GPT models performed particularly well, with the dialogue-enhanced LlaMA Chat still underperforming compared with LlaMA. This suggests that LLMs in the field of depression detection are subject to certain artificial limitations, impacting their effectiveness in this specific application. The details of the template can be seen on Appendix~\ref{appendix:D}.

\vspace{-3mm}
\subsection{Lora VS P-tuning}
From our previous ablation experiments, we found that the conventional method of incorporating LoRA matrices into attention layers might not be well-suited for depression detection tasks. After experimenting with applying LoRA matrices across all layers and conducting a hyperparameter search, we observed that LoRA, in this context, achieved results similar to those of P-tuning. Furthermore, in our use of LoRA for classification tasks, we tested a variety of manually crafted templates. However, none were as effective as using no task-specific prompt template. We believe this occurs because when we explicitly inform the LLMs that the task involves depression detection, the model tends to avoid responses that could pose ethical risks.

\vspace{-2mm}
\section{Conclusion}
\vspace{-2mm}
This paper introduces an efficient approach for depression detection using acoustic landmarks and LLMs. This approach is not only valuable for the detection of depression but also represents a new perspective in enhancing the ability of LLMs to comprehend speech signals. Furthermore, we are the \textbf{first to research multimodal depression detection using LLMs}. We establish a new benchmark with a SOTA F1-score of 0.84 through ensemble learning. Additionally, we evaluated various PEFT methods and discovered that applying Lora across all layers yields identical outcomes for both P-tuning and Lora in depression detection. Our analysis further reveals how LLMs process speech landmarks, guiding future research in this domain.

\clearpage
\section*{Limitations}
In addition, The study is confined to the DAIC-WOZ dataset, which is currently the most commonly used and only publicly available dataset in the field of multimodal depression recognition, particularly in the area of speech. The difficulty in acquiring data due to numerous privacy concerns surrounding depression datasets is acknowledged. Despite the limitations of focusing on this single dataset, it aligns with traditional research methodologies in this domain, as previous studies have predominantly relied on it.

\section*{Ethics Statement}
The DAIC-WOZ datasets are publicly available benchmarks and have been automatically de-identifed to protect patient privacy. Although our model improves the factual accuracy of generated reports, its performance still lags behind the needs of practical deployment. The outputs of our model may contain false observations and diagnoses due to systematic biases. In this regard, we strongly urge the users to examine the generated output in real-world applications cautiously.

\section*{Acknowledgement}
This work was supported by Australian Research Council Discovery Project DP230101184. 

\bibliography{reference}
\bibliographystyle{acl_natbib}
\appendix
\section{Details of Landmark Detection}
\label{appendix:A}

\subsection{General Processing Details}
Given a discrete time series signal $x[n]$, the process of peak detection consists of several pre-processing steps, followed by the identification of significant peaks. The steps are as follows:

\subsection*{Six Frequency Bands}
The following table describes the six frequency bands we used in our algorithm.
\begin{table}[htbp]
  \centering
  \caption{Frequency Bands}
  \begin{tabular}{cc}
    \toprule
    {Band} & {Frequency Range (kHz)} \\
    \midrule
    1 & 0.0--0.4 \\
    2 & 0.8--1.5 \\
    3 & 1.2--2.0 \\
    4 & 2.0--3.5 \\
    5 & 3.5--5.0 \\
    6 & 5.0--8.0 \\
    \bottomrule
  \end{tabular}
  \label{tab:frequency_bands}
\end{table}

\subsection*{Coarse Smoothing}
The signal is first subjected to a coarse smoothing operation to reduce noise and highlight broader trends. This is achieved by applying a centered moving average with a window size of $cp\_sm$:
\begin{equation}
  \scalebox{0.7}{$L^{(cp)}_{b}[n] = 10 \cdot \log_{10}\left( \frac{1}{2N_{cp}+1} \sum_{k=-N_{cp}}^{N_{cp}} E_{b}[n+k] \right)$}
\end{equation}

where $E_{b}[n]$ is the energy in the $b^{th}$ frequency band at time $n$, and $N_{cp}$ is half the size of the coarse smoothing window.

\subsection*{Coarse Differentiation}
The smoothed signal undergoes differentiation to identify regions of rapid change, which could indicate potential peaks. The differentiation is centered on mitigating delay:
\begin{equation}
  D^{(cp)}_{b}[n] = L^{(cp)}_{b}[n+cp\_dt] - L^{(cp)}_{b}[n],
\end{equation}
followed by a shift to center the result:
\begin{equation}
  D^{(cp)}_{b}[n] \leftarrow D^{(cp)}_{b}[n - \lfloor cp\_dt/2 \rfloor].
\end{equation}

\subsection*{Fine Smoothing}
A finer smoothing operation is applied to the original signal to preserve more detail, with a window size of $fp\_sm$:
\begin{equation}
  \scalebox{0.7}{$L^{(fp)}_{b}[n] = 10 \cdot \log_{10}\left( \frac{1}{2N_{fp}+1} \sum_{k=-N_{fp}}^{N_{fp}} E_{b}[n+k] \right)$}
\end{equation}

where $N_{fp}$ is half the size of the fine smoothing window.

\subsection*{Fine Differentiation}
As with coarse differentiation, the finely smoothed signal is differentiated:
\begin{equation}
  D^{(fp)}_{b}[n] = L^{(fp)}_{b}[n+fp\_dt] - L^{(fp)}_{b}[n],
\end{equation}
and then centered:
\begin{equation}
  D^{(fp)}_{b}[n] \leftarrow D^{(fp)}_{b}[n - \lfloor fp\_dt/2 \rfloor].
\end{equation}

\subsection*{Peak Detection}
After pre-processing, peaks are identified using the conditions specified earlier, considering factors such as prominence, height, and minimum distance between peaks.

Given a signal sequence $x[n]$, the peak detection process can be mathematically described as follows:

A data point $x[n]$ is considered a local maximum if it satisfies the following condition:
\begin{equation}
  x[n] > x[n-1] \quad \text{and} \quad x[n] > x[n+1].
\end{equation}

If a height threshold $h$ is specified, $x[i]$ is recognized as a peak only if:
\begin{equation}
  x[i] > h.
\end{equation}

The prominence $P$ of a peak at $x[i]$ is defined as the vertical distance between the peak and its lowest contour line:
\begin{equation}
  P = x[i] - \max(v_l, v_r),
\end{equation}
where $v_l$ and $v_r$ are the lowest points on either side of $x[i]$, before reaching a higher point. A peak is considered significant if its prominence exceeds a predefined threshold.

The width $W$ of a peak is measured at a vertical distance $P$ from its highest point. Points $x[l]$ and $x[r]$, where $l < i < r$, are the positions at which the signal drops below the threshold defined by the prominence:
\begin{equation}
  x[l] < x[i] - P \quad \text{and} \quad x[r] < x[i] - P,
\end{equation}
and the width $W$ is the distance between $x[l]$ and $x[r]$.

If a minimum peak separation distance $D$ is defined, then for any two peaks $x[i]$ and $x[j]$, the condition must be met:
\begin{equation}
  |i - j| > D.
\end{equation}

These conditions are used to identify peaks in the signal that are not only local maxima but also exceed certain amplitude and prominence thresholds, ensuring the detected peaks are significant in the context of the signal.

\subsection{Details of Specific Landmark Detection} 
\textbf{g landmark} When both the coarse and fine filters exhibit a peak in band 1, it is identified as a 'g' landmark. 

\textbf{b landmark} In an unvoiced segment (not between +g and the next -g), if at least three out of five frequency bands demonstrate simultaneous power increases of no less than 6 dB in both coarse and fine filters, a specific condition or criterion is met.

\textbf{s landmark} In an unvoiced segment (between +g and the next -g), if at least three out of five frequency bands demonstrate simultaneous power increases of no less than 6 dB in both coarse and fine filters, a specific condition or criterion is met.

\textbf{f+} and \textbf{v+ landmarks} involves detecting a 6 dB power increase in at least three high-frequency bands (4, 5, 6), and a power decrease in low-frequency bands (2, 3). For \textbf{f-} and \textbf{v-}, the criteria are reversed: a 6 dB power decrease in the same high-frequency bands and a power increase in the low-frequency bands.The distinguishing factor here is that frication landmarks are detected within unvoiced segments (b landmark), while voiced frication landmarks are sought in voiced segments (s landmark).

\textbf{p landmark}, p landmark extraction can be divided into several steps.
\newline
1. \textbf{Frame Segmentation}: \\
Let the audio signal be \( Y(t) \). \\
Define the frame length \( N \) and frame shift \( \Delta \). \\
For the \( i \)-th frame, we consider the segment \( Y[i \cdot \Delta : i \cdot \Delta + N] \).
\newline
2. \textbf{Autocorrelation Calculation}: \\
For each frame \( Y_i \), calculate the autocorrelation function \( R_{xx}(k) \): \\
\[ R_{xx}(k) = \frac{1}{N-k} \sum_{n=0}^{N-k-1} Y_i(n) \cdot Y_i(n+k). \]
\newline
3. \textbf{Energy Function Calculation}: \\
Compute the energy function \( E_f \) for each frame: \\
\[ E_f(i) = \frac{1}{N} \sum_{k=0}^{N-1} R_{xx}(k)^2. \]
\newline
4. \textbf{Upsampling}: \\
Upsample the energy function \( E_f \) to match the length of the original signal.
\newline
5. \textbf{Smoothing}: \\
Apply smoothing(As defined in the previous section) to the upsampled energy function.
\newline
6. \textbf{Binarization}: \\
Define a threshold \( \theta \), and convert the smoothed energy function into a binary signal.
\newline
7. \textbf{Jump Detection}: \\
Detect positive and negative jumps in the binary signal.
\newline
8. \textbf{P Landmark Index and Time Determination}: \\
Record the positions of jumps, which are the indices of P landmarks. \\
Convert these indices into time points to determine the P landmarks.

\begin{algorithm}[t]
\footnotesize  
\caption{Sub-dialogue shuffling}
\begin{algorithmic}[1]
\State $N^+ \gets$ Number of positive samples in the training set
\State $N^- \gets$ Number of negative samples in the training set
\State $M \gets$ Set number of sub-dialogues for each positive sample $M^+$
\State $M^* \gets N^- / N^+$
\State Set $\varepsilon_l, \varepsilon_h$ satisfying $0 < \varepsilon_l < \varepsilon_h \leq 1$
\For{Dialogue $X^{(n)}$ $n = 1$ \textbf{to} $N$}
    \State $T \gets \text{len}(x^{(n)})$
    \If{$x^{(n)}$ is positive}
        \State $M \gets M^+$
    \Else
        \State $M \gets M^-$
    \EndIf
    \For{Sub-dialogue $X^{(n)m}$ $m = 1$ \textbf{to} $M$}
        \State Sample $\varepsilon$ uniformly from $(\varepsilon_l, \varepsilon_h)$
        \State $d \gets \varepsilon T - 1$
        \State Sample $s$ randomly from range $(0, T-d)$
        \State $e \gets s + d$
        \State $X^{(n)m} \gets x_{s:e}^{(n)}$
    \EndFor
\EndFor
\end{algorithmic}
\end{algorithm}

\section{Details of Data Augmentation}

\label{appendix:B}

The training set was expanded by shuffling sub-dialogues, selecting portions $x_{s:e}$ from each full dialogue $x_{1:T}$, with $s$ and $e$ as random start and end indices. The algorithm outlines this process. Initially, it counts the positive and negative samples, setting $M^+$ as the target number of sub-dialogues for each positive dialogue (Algorithm 1, lines 1-3). To balance augmentation, $M^-$ is calculated using $N^+$, $N^-$, and $M^+$ (line 4). For both positive and negative dialogues, corresponding $M^+$ and $M^-$ sub-dialogues are generated (lines 8-12). The sub-dialogue length, $d$, is set within the range defined by $\varepsilon_l$ and $\varepsilon_h$, chosen randomly (lines 14-15). The start index $s$ is randomly selected within its range, and the end index $e$ is determined accordingly (lines 16-18)~\cite{wu2023self}.

\section{Sample of Hint Cross-modal Instruction Fine Tuning}
\label{appendix:C}

\textbf{Depression Example}
\begin{lstlisting}[basicstyle=\ttfamily\footnotesize, breaklines=true]
Below are the speech transcripts from a person with depression.
Please try to predict the concatenated acoustic landmarks
corresponding to these transcripts.

### Transcript:
{transcript}

### Acoustic Landmark:
{landmark}
\end{lstlisting}

\textbf{Healthy Example}
\begin{lstlisting}[basicstyle=\ttfamily\footnotesize, breaklines=true]
Below are the speech transcripts from a healthy person.
Please try to predict the concatenated acoustic landmarks
corresponding to these transcripts.

### Transcript:
{transcript}

### Acoustic Landmark:
{landmark}
\end{lstlisting}

\section{Sample of Instruction Fine-Tuning for Depression Detection}
\label{appendix:D}

\textbf{Text Only}
\begin{lstlisting}[basicstyle=\ttfamily\footnotesize, breaklines=true]
"Categorize these dialogues as either depression or healthy based on its transcripts.

### transcript:{transcript}

### Response:"
\end{lstlisting}

\textbf{Landmark Only}
\begin{lstlisting}[basicstyle=\ttfamily\footnotesize, breaklines=true]
"Categorize these dialogues as either depression or healthy based on its acoustic landmarks.

### acoustic landmarks:{landmarks}

### Response:"
\end{lstlisting}

\textbf{MultiModal}
\begin{lstlisting}[basicstyle=\ttfamily\footnotesize, breaklines=true]
"Categorize these dialogues as either depression or healthy based on its transcripts and acoustic landmarks.

### Transcript:{transcript}

### Acoustic Landmark:{landmarks}

### Response:\n"
\end{lstlisting}

\end{document}